\documentclass[sigconf,screen=true,authorversion]{acmart}

\AtBeginDocument{%
  }

\copyrightyear{2024}
\acmYear{2024}
\setcopyright{acmlicensed}\acmConference[SIGIR-AP '24]{Proceedings of the 2024 Annual International ACM SIGIR Conference on Research and Development in Information Retrieval in the Asia Pacific Region}{December 9--12, 2024}{Tokyo, Japan}
\acmBooktitle{Proceedings of the 2024 Annual International ACM SIGIR Conference on Research and Development in Information Retrieval in the Asia Pacific Region (SIGIR-AP '24), December 9--12, 2024, Tokyo, Japan}
\acmDOI{10.1145/3673791.3698423}
\acmISBN{979-8-4007-0724-7/24/12}

\usepackage{graphicx}
\usepackage{booktabs}
\usepackage[edges]{forest}
\usepackage{acronym} 
\usepackage{tcolorbox}
\usepackage{multicol}
\usepackage{enumitem}
\usepackage{xcolor}
\usepackage{fontawesome}
\usepackage{balance}

\newcommand\crule[3][black]{\textcolor{#1}{\rule{#2}{#3}}}

\begin{document}

\acrodef{IR}[IR]{information retrieval}
\acrodef{LLM}[LLM]{Large Language Model}
\acrodefplural{LLM}[LLMs]{Large Language Models}
\acrodef{COT}[COT]{chain-of-thought}
\acrodef{UQV}[UQV]{user query variants}
\acrodef{RRF}[RRF]{Reciprocal Rank Fusion}
\acrodef{RAG}[RAG]{Retrieval Augmented Generation}

\title[Data Fusion of Synthetic Query Variants With Generative LLMs]{Data Fusion of Synthetic Query Variants \\ With Generative Large Language Models}

\author{Timo Breuer}
\orcid{0000-0002-1765-2449}
\affiliation{
  \institution{TH Köln --- University of Applied Sciences}
  \city{Cologne}
  \country{Germany}
}
\email{timobreuer@acm.org}

\renewcommand{\shortauthors}{Timo Breuer}

\begin{abstract}

  Considering query variance in \ac{IR} experiments is beneficial for retrieval effectiveness. Especially ranking ensembles based on different topically related queries retrieve better results than rankings based on a single query alone. Recently, generative instruction-tuned \acp{LLM} improved on a variety of different tasks in capturing human language. To this end, this work explores the feasibility of using synthetic query variants generated by instruction-tuned \acp{LLM} in data fusion experiments. More specifically, we introduce a lightweight, unsupervised, and cost-efficient approach that exploits principled prompting and data fusion techniques. In our experiments, \acp{LLM} produce more effective queries when provided with additional context information on the topic. Furthermore, our analysis based on four TREC newswire benchmarks shows that data fusion based on synthetic query variants is significantly better than baselines with single queries and also outperforms pseudo-relevance feedback methods. We publicly share the code and query datasets with the community as resources for follow-up studies. 
  
  \vspace{2.5em}
  \hspace{4em}
  \parbox[c]{\columnwidth}
  {
      \vspace{1.4em}
      \faGithub \ \href{https://github.com/breuert/sigirap24}{\nolinkurl{https://github.com/breuert/sigirap24}}
  }
  \vspace{2em}

\end{abstract}


\begin{teaserfigure}
  \centering
  \includegraphics[width=.9\textwidth]{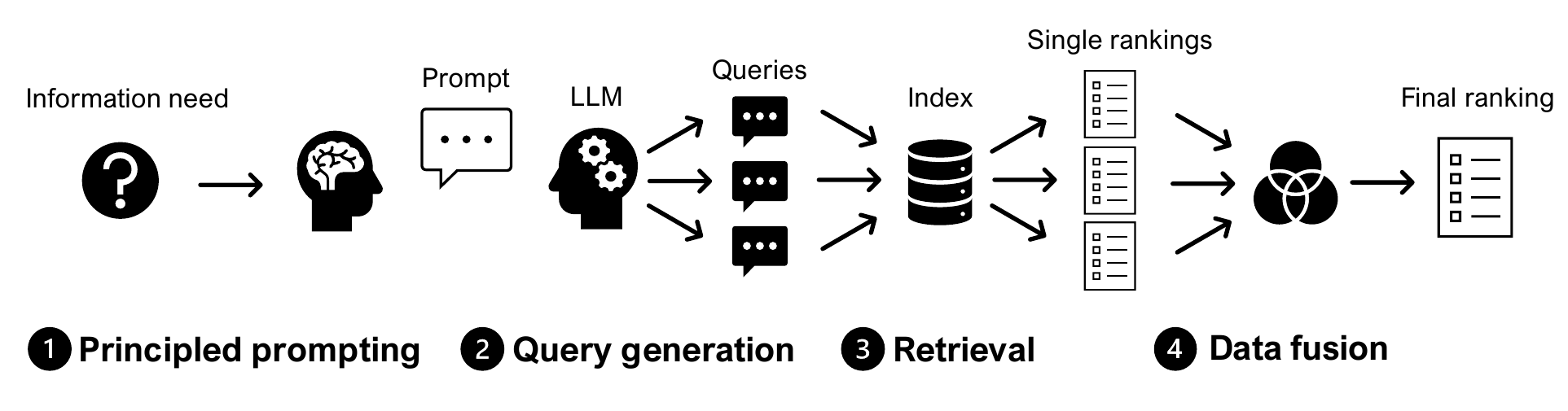}
  \caption{Data fusion of synthetic query variants includes 1) principled prompting, 2) query variant generation with instruction-tuned Large Language Models, 3) retrieval of single rankings, and 4) data fusion.}
  \label{fig:methodology}
\end{teaserfigure}

\maketitle

\section{Introduction}

Test collection-based evaluations of \ac{IR} experiments are typically done in accordance with the Cranfield paradigm. Usually, these experiments are implemented with topics that define the users' information need in a structured way~\cite{DBLP:journals/ftir/Sanderson10}. In principle, there are different ways to generate queries from topic files. However, the de facto standard in most \ac{IR} evaluations is the use of the topic's title as the query string. While this approach allows better comparability between different systems, it completely neglects query variability and also the system behavior under the consideration of different topically or semantically related queries~\cite{DBLP:conf/sigir/BaileyMST15}.

Including multiple query representations in the retrieval process bears a lot of potential to improve retrieval effectiveness. Empirically, the improvements by data fusion algorithms were shown in several works. However, earlier work either required crowdsourcing efforts to obtain real query variants from users \cite{DBLP:conf/sigir/BaileyMST16,DBLP:conf/adcs/BenhamC17}, which is costly and time-consuming, or generated synthetic query variants for which additional query modeling methods were necessary~\cite{DBLP:conf/cikm/ChakrabortyGC20,DBLP:conf/ecir/BreuerFS22}.

Recently, generative instruction-tuned \acp{LLM} substantially improved on various language modeling tasks. \textit{Zero shot, few shot}, or \textit{in-context learning}~\cite{DBLP:conf/nips/BrownMRSKDNSSAA20} does not require additional task-specific fine-tuning but instead provides the task-agnostic language model with context information based on precise instructions and by learning from examples in the prompt, allowing the language model to grasp the task. These advances offer promising solutions to reduce efforts when in need of obtaining query variants.

To this end, this work evaluates if synthetic query variants based on generative \acp{LLM} can be used to improve the retrieval effectiveness in data fusion experiments. Figure~\ref{fig:methodology} provides a high-level illustration of our proposed methodology. Given the information need that is described by the topic file, we construct a topic-specific prompt that is given to the \ac{LLM}, and the generated queries are then used to retrieve documents from the index, yielding multiple single rankings based on different query variants. Afterward, the single rankings are combined with data fusion techniques, resulting in the final ranking. 
Our results suggest that fused rankings of synthetic query variants based on \acp{LLM} and principled prompting can result in viable results that, in combination, yield better effectiveness than the same retrieval method would yield with a single query.

While this work focuses on evaluating the retrieval effectiveness of the fused rankings, the proposed method could also be understood as a component of a \ac{RAG} system where the ranking is an intermediate used to improve the quality of system outputs in a conversation with an agent. In such a conversational search setting, it is also feasible to obtain a more precise understanding of the user's information need, similar to the topic file, making this approach a promising candidate for real-world applications, which we discuss later in the text. In sum, our contributions are threefold and include the following:

\begin{itemize}
  \item An \textbf{unsupervised and lightweight retrieval approach} based on data fusion of synthetic query variants,
  \item \textbf{experimental evaluations} based on four different newswire datasets, including different prompting strategies and different numbers of fused query variants,
  \item \textbf{new datasets of query variants} for TREC test collections of the respective tracks and \textbf{open-source code of the experimental setup} for better reproducibility and follow-up studies.\footnote{\faGithub \ \url{https://github.com/breuert/sigirap24}}
\end{itemize}

\section{Related Work}
\label{sec:related_work}

Recently, Alaofi et al.~\cite{DBLP:conf/sigir/AlaofiGMSSSSW22} highlighted that the source of query variation is little studied. Currently, there is no cohesive framework that explains how users come up with query formulations. Bailey et al.~\cite{DBLP:conf/sigir/BaileyMST15} pointed out the importance of user variance and concluded that query variance leads to similar variability in retrieval effectiveness as topical or system variance. In this regard, Moffat et al.~\cite{DBLP:conf/cikm/MoffatSTB15} showed that query variability has a similar effect on the pool size as system variability. One of the earlier works about data fusion by Belkin et al.~\cite{DBLP:conf/sigir/BelkinCCC93,DBLP:journals/ipm/BelkinKFS95} showed that fusing rankings based on the same retrieval method and different queries can be more effective than fusing rankings based on different retrieval methods that used a single query variant. Similarly, data fusion can also be understood as an operationalization of polyrepresentation~\cite{DBLP:journals/jd/Ingwersen96,DBLP:journals/jasis/LarsenIL09}. Benham et al.~\cite{DBLP:conf/trec/BenhamGMDCSMC17,DBLP:conf/trec/BenhamGML0SCM18,DBLP:conf/desires/BenhamCGLM18,DBLP:journals/tois/BenhamMMC19,DBLP:conf/adcs/BenhamC17} gave empirical evidence to the effectiveness of fusing rankings based on user query variants in several works, primarily showing improvements in the recall rates.

Baskaya et al.~\cite{DBLP:conf/cikm/BaskayaKJ13} proposed several principled formulation strategies to generate synthetic query variants from topic texts, i.e., given the topic text, they apply term variations or extensions for sequential query formulations. Azzopardi et al.~\cite{DBLP:conf/sigir/AzzopardiRB07} introduced a testbed for simulated queries of known items. The corresponding known-item queries were built as proposed by Jordan et al.~\cite{DBLP:conf/jcdl/JordanWG06}. The underlying language model is based on the term distributions in relevant documents and requires apriori knowledge about relevant documents. Benham and Culpepper~\cite{DBLP:conf/desires/BenhamCGLM18} generated synthetic queries with expansion techniques based on large external corpora, and used them to fuse rankings from a target collection. There exist more recent transformer-based query generation methods like Doc2Query, as proposed by Nogueira et al.~\cite{DBLP:journals/corr/abs-1904-08375}, where a sequence-to-sequence transformer is used to generate a query from document terms. While this approach has been primarily used to augment documents during indexing~\cite{DBLP:conf/ecir/GospodinovMM23}, the generated queries could also be used for data fusion experiments if suitable and topically related documents of an external corpus are available for the query generation. Penha et al.~\cite{DBLP:conf/ecir/PenhaCH22} proposed a taxonomy of different user query variants for which they implement different query generators. For instance, they fine-tuned a pretrained T5 model to generate title queries from descriptions of topic texts. 

As an alternative to fine-tuning a task-specific \ac{LLM}, \textit{in-context learning}~\cite{DBLP:conf/nips/BrownMRSKDNSSAA20} does not require additional fine-tuning for many tasks. Once the model is fine-tuned for instructions, it can be applied to many specific downstream tasks based on the prompted instructions. For instance, \textit{chain-of-thought prompting}~\cite{DBLP:journals/corr/abs-2201-11903} is a simple but effective technique that allows \acp{LLM} to pick up task-specific reasoning by providing examples in the prompt. Recently, several prompt-based query generation techniques were proposed. Alaofi et al.~\cite{DBLP:conf/sigir/AlaofiGSS023} provided a first experimental evaluation of prompt-based \ac{LLM}-generated query variants in reference to the UQV100 dataset provided by Bailey et al.~\cite{DBLP:conf/sigir/BaileyMST16}. Jagerman et al.~\cite{DBLP:journals/corr/abs-2305-03653} proposed a query expansion technique~ based on prompting the \ac{LLM} to answer a query, which is then expanded by the generated output to retrieve a ranking with a lexical retrieval method like BM25. Dai et al.~\cite{DBLP:conf/iclr/DaiZMLNLBGHC23} propose a few-shot prompting technique that makes it possible to generate effective queries by providing the \ac{LLM} with a sample of queries.

\section{Methodology}
\label{sec:methodology}

This section provides details about our methodology that is illustrated from a high-level perspective in Figure~\ref{fig:methodology}. More specifically, we describe the approach of prompting the \ac{LLM} and generating queries (cf. \ref{sec:prompting}), review the data fusion method (cf. \ref{subsec:data.fusion}), and provide an overview of the experimental setup (cf. \ref{subsec:datasets_implementation}).

\subsection{Principled Prompting and Query Generation}
\label{sec:prompting}

The prompts sent to the instruction-tuned \ac{LLM} are constructed in principled ways by following common practices of prompting strategies. We specify the role of the \ac{LLM} and the format of the outputs. Then, we augment the prompt with the contents of the topic file and compose the prompt as shown below.

\begin{tcolorbox}[title=\textbf{Prompt template for strategies P-1, P-2, and P-3},size=small,fontupper=\small, fontlower=\small]
  \texttt{\textbf{You are a generator of search query variants.}} \\ 
  
  \texttt{\textcolor{cyan}{Generate one hundred \textcolor{cyan}{keyword} queries about \textbf{<title>}}.} \\ 
  
  \texttt{\textcolor{purple}{\textbf{<description>} \textbf{<narrative>}}} \\
  
 \texttt{\textcolor{teal}{Example queries for the topic about \textbf{<example title>} include \textbf{<1st query example>}, \textbf{<2nd query example>}, \textbf{<3rd query example>} ...}} \\ 
  
  \texttt{\textbf{Your reply is a numbered list of search queries.}}
\end{tcolorbox}

The first prompting strategy \textbf{P-1} \crule[black]{.3cm}{.3cm}\crule[cyan]{.3cm}{.3cm}\crule[black]{.3cm}{.3cm} includes the topic's title as topical context information besides the prompt to generate one hundred queries, spanned by the role definition and the specification of the output format. In comparison, the second strategy \textbf{P-2} \crule[black]{.3cm}{.3cm}\crule[cyan]{.3cm}{.3cm}\crule[purple]{.3cm}{.3cm}\crule[black]{.3cm}{.3cm} adds the topic's description and narrative as context information. Wei et al.~\cite{DBLP:journals/corr/abs-2201-11903} showed that \textit{chain-of-thought prompting}  can lead to significant performance improvements on tasks such as arithmetic, commonsense, and symbolic reasoning. In this regard, we extend the prompt with examples of query reformulations. To avoid any overlap of topics with our test collections, we use examples of the dataset provided by Bailey et al.~\cite{DBLP:conf/sigir/BaileyMST16}. More specifically, the query examples were picked from the topics 201, 251, 263, 266, and 296 of the 2013 and 2014 TREC Web tracks \cite{DBLP:conf/trec/Collins-Thompson13,DBLP:conf/trec/Collins-Thompson14} for the third prompting strategy \textbf{P-3} \crule[black]{.3cm}{.3cm}\crule[cyan]{.3cm}{.3cm}\crule[teal]{.3cm}{.3cm}\crule[black]{.3cm}{.3cm}. We use this dataset of query variants for the query examples to avoid query data leakage. The particular topics were randomly chosen from a preselection of topics with a sufficient number of query variants.

In order to generate the query variants, we rely on OpenAI's \texttt{gpt-4o}.\footnote{\url{https://platform.openai.com/docs/models/gpt-4o}} At the time of our experiments, this identifier points to the model version \texttt{gpt-4o-2024-05-13} with training data up to October 2023. Besides, we make the API calls with a fixed seed and a temperature parameter of $0.0$. Since the model reliably returns numbered lists, the parsing of the generated outputs is fairly simple and straightforward. To illustrate the costs of query generation, let us consider all of the 400 topics of our test collections. For each topic, we generate one hundred queries. Generating for each of the three different prompts 40,000 queries translated into approx. 1.5 million generated tokens. In sum, we curated four datasets of query variants for TREC test collections covering three prompting strategies with over 120,000 query strings for approximately \$22.50.\footnote{Based on \textit{US\$15.00 / 1M tokens} as of July 2024.}

\subsection{Data Fusion}
\label{subsec:data.fusion}

Data fusion methods can be categorized into rank- or score-based algorithms. In the data fusion experiments, we rely on \ac{RRF}~\cite{DBLP:conf/sigir/CormackCB09} --- a rank-based fusion method --- that proved to be effective and robust. It is defined as follows:
\begin{equation}
  R R F \operatorname{score}(d \in D)=\sum_{r \in R} \frac{1}{k+r(d)}
  \label{eq:rrf}
\end{equation}
where $R$ denotes a set of rankings, $r(d)$ denotes the rank of a documents $d$ from a set of documents $D$ in a particular ranking $r$, and $k$ is a free parameter set to $60$ to comply with earlier work. Even though it is a straightforward method, it demonstrated its effectiveness over \textit{learning to rank} approaches and other data fusion methods. We note that there are other probabilistic data fusion methods~\cite{DBLP:conf/sigir/AslamM01,DBLP:conf/sigir/LillisTCD06,DBLP:conf/ecir/Shokouhi07a,DBLP:conf/cikm/WuC02} that are effective but require additional fine-tuning or optimization with training data. Opposed to these methods, \ac{RRF} can be used \textit{off-the-shelf}, which is intentional as the methodology is intended to be unsupervised. Besides, we acknowledge its widespread application and use in several studies~\cite{DBLP:conf/jcdl/BreuerKST23,DBLP:conf/cikm/BassaniR22,DBLP:conf/sigir/VoskaridesLRKR20,DBLP:conf/sigir/SalemiKZ24,DBLP:journals/tois/LinYNTWL21,DBLP:conf/sigir/KhramtsovaZBZ24}, also being implemented in the popular open-source software toolkit Elasticsearch.\footnote{\url{https://www.elastic.co/guide/en/elasticsearch/reference/8.15/rrf.html}} % archived at: https://archive.ph/QTPZQ

\subsection{Datasets and Implementation Details}
\label{subsec:datasets_implementation}

Our experiments use four TREC newswire test collections including TREC Disks 4 \& 5 (minus congressional records)\footnote{\url{https://trec.nist.gov/data/cd45/}} used as part of TREC Robust 2004 (\textbf{Robust04})~\cite{DBLP:conf/trec/Voorhees04b}, the AQUAINT Corpus of English News Text\footnote{\url{https://catalog.ldc.upenn.edu/LDC2002T31}} used as part of TREC Robust 2005 (\textbf{Robust05})~\cite{DBLP:journals/sigir/Voorhees06}, the New York Times Annotated Corpus\footnote{\url{https://catalog.ldc.upenn.edu/LDC2008T19}} used as part of TREC Common Core 2017 (\textbf{Core17})~\cite{DBLP:conf/trec/AllanHKLGV17}, and the TREC Washington Post Corpus\footnote{\url{https://trec.nist.gov/data/wapost/}} used as part of TREC Common Core 2018 (\textbf{Core18})~\cite{DBLP:conf/trec/2018}.

Our implementations are based on a set of state-of-the-art retrieval and evaluation toolkits including Pyterrier~\cite{DBLP:conf/cikm/MacdonaldTMO21}, building upon the Terrier toolkit~\cite{DBLP:conf/ecir/OunisAPHMJ05}, \texttt{ir\_datasets}~\cite{DBLP:conf/sigir/MacAvaneyYFDCG21}, and \texttt{ranx}~\cite{DBLP:conf/ecir/Bassani22,DBLP:conf/cikm/BassaniR22}. For the indexing, we use \texttt{ir\_datasets} as the toolkit provides bindings for all four datasets. For the retrieval, we rely on BM25 with default settings as implemented in Pyterrier/Terrier. All of the different data fusion methods are based on \texttt{ranx.fuse}~\cite{DBLP:conf/cikm/BassaniR22} which is an extension to \texttt{ranx}~\cite{DBLP:conf/ecir/Bassani22}. Likewise, we use \texttt{ranx} to evaluate the experiments. 

\section{Experimental Results}
\label{sec:experimental_results}

In our experimental evaluations, we consider two baseline retrieval methods, namely BM25 and BM25 + RM3. The BM25 baseline rankings are retrieved with the topic's title as provided. BM25 + RM3 is a pseudo-relevance feedback method that builds upon expanding the initial query after retrieving a first ranking for determining expansion terms. After the initial ranking is retrieved with the topic's title, an additional (and final) ranking is retrieved with the initial query and expansion terms based on RM3. 

Table~\ref{tab:prompts.effectiveness} shows the retrieval effectiveness in terms of P@10, nDCG@10, Bpref, and MAP evaluated with four newswire datasets. For all fused rankings, we use a set of ten rankings based on the first ten synthetic queries that are combined with \ac{RRF}. In principle, it is possible to generate an arbitrary number of queries. However, in practice, efficiency tradeoffs have to be made when generating queries with an \ac{LLM} on the fly. Thus, we consider ten queries as a reasonable number for evaluating the effectiveness benefits without imposing too much latency caused by the query generation.

\begin{table}[!t]
    \centering
    \caption{Comparison of fused rankings with synthetic queries based on different prompting strategies.
    Superscripts denote significant differences in paired Student's t-test (with Bonferroni correction applied).
    The best results are bold-faced.
    }
    \resizebox{0.465\textwidth}{!}{
    \begin{tabular}{c|l|c|c|c|c}
    \toprule
    \textbf{\#}
    & \textbf{Prompt}
    & \textbf{P@10}
    & \textbf{NDCG@10}
    & \textbf{BPref}  
    & \textbf{MAP} \\
    \midrule
    \multicolumn{6}{c}{\textbf{Core17}} \\
    \midrule
    a &
    BM25 &
    0.458\hphantom{$^{bcde}$} &
    0.372\hphantom{$^{bcde}$} &
    0.274\hphantom{$^{bcde}$} &
    0.199\hphantom{$^{bcde}$} \\
    b &
    BM25 + RM3 &
    0.534$^{a}$\hphantom{$^{cde}$} &
    0.404\hphantom{$^{acde}$} &
    0.317$^{a}$\hphantom{$^{cde}$} &
    0.246$^{a}$\hphantom{$^{cde}$} \\
    c &
    P-1 \crule[black]{.3cm}{.3cm}\crule[cyan]{.3cm}{.3cm}\crule[black]{.3cm}{.3cm} &
    0.526$^{a}$\hphantom{$^{bde}$} &
    0.426$^{a}$\hphantom{$^{bde}$} &
    0.355$^{ab}$\hphantom{$^{de}$} &
    0.240$^{a}$\hphantom{$^{bde}$} \\
    d &
    P-2 \crule[black]{.3cm}{.3cm}\crule[cyan]{.3cm}{.3cm}\crule[purple]{.3cm}{.3cm}\crule[black]{.3cm}{.3cm} &
    \textbf{0.618}$^{ac}$\hphantom{$^{be}$} &
    \textbf{0.522}$^{abc}$\hphantom{$^{e}$} &
    \textbf{0.416}$^{abce}$\hphantom{} &
    \textbf{0.299}$^{abce}$\hphantom{} \\
    e &
    P-3 \crule[black]{.3cm}{.3cm}\crule[cyan]{.3cm}{.3cm}\crule[teal]{.3cm}{.3cm}\crule[black]{.3cm}{.3cm} &
    0.570$^{a}$\hphantom{$^{bcd}$} &
    0.453$^{a}$\hphantom{$^{bcd}$} &
    0.376$^{ab}$\hphantom{$^{cd}$} &
    0.255$^{a}$\hphantom{$^{bcd}$} \\
    \midrule
    \multicolumn{6}{c}{\textbf{Core18}} \\ 
    \midrule
    a &
    BM25 &
    0.426\hphantom{$^{bcde}$} &
    0.389\hphantom{$^{bcde}$} &
    0.253\hphantom{$^{bcde}$} &
    0.191\hphantom{$^{bcde}$} \\
    b &
    BM25 + RM3 &
    0.448\hphantom{$^{acde}$} &
    0.396\hphantom{$^{acde}$} &
    0.282$^{a}$\hphantom{$^{cde}$} &
    0.229$^{a}$\hphantom{$^{cde}$} \\
    c &
    P-1 \crule[black]{.3cm}{.3cm}\crule[cyan]{.3cm}{.3cm}\crule[black]{.3cm}{.3cm} &
    0.452\hphantom{$^{abde}$} &
    0.433\hphantom{$^{abde}$} &
    0.294\hphantom{$^{abde}$} &
    0.233$^{a}$\hphantom{$^{bde}$} \\
    d &
    P-2 \crule[black]{.3cm}{.3cm}\crule[cyan]{.3cm}{.3cm}\crule[purple]{.3cm}{.3cm}\crule[black]{.3cm}{.3cm} &
    \textbf{0.532}$^{ace}$\hphantom{$^{b}$} &
    \textbf{0.497}$^{abce}$\hphantom{} &
    \textbf{0.339}$^{abc}$\hphantom{$^{e}$} &
    \textbf{0.270}$^{ac}$\hphantom{$^{be}$} \\
    e &
    P-3 \crule[black]{.3cm}{.3cm}\crule[cyan]{.3cm}{.3cm}\crule[teal]{.3cm}{.3cm}\crule[black]{.3cm}{.3cm} &
    0.440\hphantom{$^{abcd}$} &
    0.420\hphantom{$^{abcd}$} &
    0.312$^{a}$\hphantom{$^{bcd}$} &
    0.240$^{a}$\hphantom{$^{bcd}$} \\
    \midrule
    \multicolumn{6}{c}{\textbf{Robust04}} \\ 
    \midrule
    a &
    BM25 &
    0.410\hphantom{$^{bcde}$} &
    0.421\hphantom{$^{bcde}$} &
    0.241\hphantom{$^{bcde}$} &
    0.228\hphantom{$^{bcde}$} \\
    b &
    BM25 + RM3 &
    0.443$^{a}$\hphantom{$^{cde}$} &
    0.443$^{a}$\hphantom{$^{cde}$} &
    0.268$^{a}$\hphantom{$^{cde}$} &
    0.262$^{a}$\hphantom{$^{cde}$} \\
    c &
    P-1 \crule[black]{.3cm}{.3cm}\crule[cyan]{.3cm}{.3cm}\crule[black]{.3cm}{.3cm} &
    0.455$^{a}$\hphantom{$^{bde}$} &
    0.471$^{a}$\hphantom{$^{bde}$} &
    0.269$^{a}$\hphantom{$^{bde}$} &
    0.256$^{a}$\hphantom{$^{bde}$} \\
    d &
    P-2 \crule[black]{.3cm}{.3cm}\crule[cyan]{.3cm}{.3cm}\crule[purple]{.3cm}{.3cm}\crule[black]{.3cm}{.3cm} &
    \textbf{0.514}$^{abce}$\hphantom{} &
    \textbf{0.535}$^{abce}$\hphantom{} &
    \textbf{0.304}$^{abce}$\hphantom{} &
    \textbf{0.295}$^{abce}$\hphantom{} \\
    e &
    P-3 \crule[black]{.3cm}{.3cm}\crule[cyan]{.3cm}{.3cm}\crule[teal]{.3cm}{.3cm}\crule[black]{.3cm}{.3cm} &
    0.459$^{a}$\hphantom{$^{bcd}$} &
    0.480$^{ab}$\hphantom{$^{cd}$} &
    0.276$^{a}$\hphantom{$^{bcd}$} &
    0.265$^{a}$\hphantom{$^{bcd}$} \\
    \midrule
    \multicolumn{6}{c}{\textbf{Robust05}} \\
    \midrule
    a &
    BM25 &
    0.352\hphantom{$^{bcde}$} &
    0.296\hphantom{$^{bcde}$} &
    0.227\hphantom{$^{bcde}$} &
    0.176\hphantom{$^{bcde}$} \\
    b &
    BM25 + RM3 &
    0.408\hphantom{$^{acde}$} &
    0.329\hphantom{$^{acde}$} &
    0.251\hphantom{$^{acde}$} &
    0.209$^{a}$\hphantom{$^{cde}$} \\
    c &
    P-1 \crule[black]{.3cm}{.3cm}\crule[cyan]{.3cm}{.3cm}\crule[black]{.3cm}{.3cm} &
    0.396\hphantom{$^{abde}$} &
    0.343\hphantom{$^{abde}$} &
    0.267$^{a}$\hphantom{$^{bde}$} &
    0.206\hphantom{$^{abde}$} \\
    d &
    P-2 \crule[black]{.3cm}{.3cm}\crule[cyan]{.3cm}{.3cm}\crule[purple]{.3cm}{.3cm}\crule[black]{.3cm}{.3cm} &
    \textbf{0.522}$^{ac}$\hphantom{$^{be}$} &
    \textbf{0.442}$^{a}$\hphantom{$^{bce}$} &
    \textbf{0.361}$^{abce}$\hphantom{} &
    \textbf{0.267}$^{ac}$\hphantom{$^{be}$} \\
    e &
    P-3 \crule[black]{.3cm}{.3cm}\crule[cyan]{.3cm}{.3cm}\crule[teal]{.3cm}{.3cm}\crule[black]{.3cm}{.3cm} &
    0.492$^{ac}$\hphantom{$^{bd}$} &
    0.404$^{ac}$\hphantom{$^{bd}$} &
    0.303$^{abc}$\hphantom{$^{d}$} &
    0.236$^{ac}$\hphantom{$^{bd}$} \\
    \bottomrule
    \end{tabular}
    }
    \label{tab:prompts.effectiveness}
\end{table}

In general, the fused rankings outperform the baselines BM25 (+ RM3) in most cases, i.e., the measures have higher but not necessarily significantly better scores. Depending on the prompting strategy, some fused rankings are more effective than others. However, we also note that for some measures and test collections, especially, P-1 is outperformed by BM25 + RM3. Still, all fused rankings perform better than BM25 rankings based on a single query.

For all four test collections, fused rankings with queries based on the prompting strategy P-2 perform best, suggesting that there is a positive effect on retrieval effectiveness when including additional information about the topic (based on the description and narrative) in the prompt given to the \ac{LLM}. In most cases, the P-2 results are significantly better than the baseline methods and the other two prompting strategies.

Regarding BPref, we see that P-2 outperforms each other method with statistical significance (except for Core18, for which P-2 has better but not significantly different effectiveness than P-3). These outcomes suggest that many unjudged documents are contained in the rankings, which is typical when using query variants. Other topically related queries can bring up potentially relevant documents that were not considered in the pooling of the test collection creation process.

In comparison, the other prompting strategies P-1 and P-3 are less effective. While both of them outperform the BM25 baseline rankings, sometimes pseudo-relevance feedback with BM25 + RM3 is more effective than fusing rankings with queries of P-1, which suggests that simply providing the title string in the prompt is not enough to guide the \ac{LLM} to produce effective and topically-related queries. More specifically, fused rankings with queries based on P-1 are less effective than BM25 + RM3 wrt. P@10 and MAP when evaluated with Core17 and Robust05. Even though less effective than P-2, the P-3 rankings are, in most cases, significantly better than BM25 and have higher scores than BM25 + RM3.

\begin{figure}[!t]
  \includegraphics[width=0.235\textwidth]{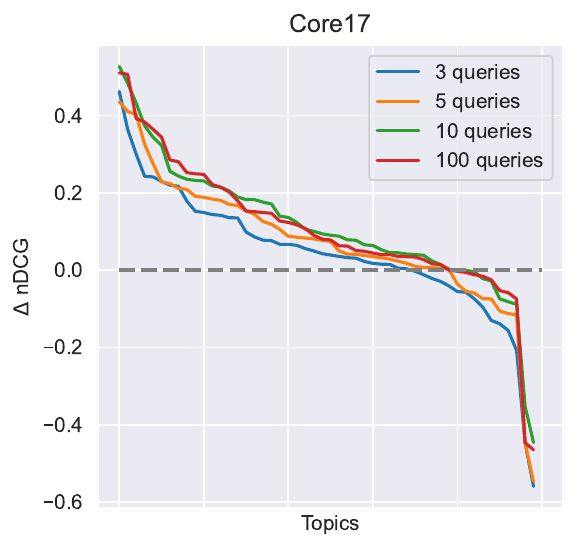}
  \includegraphics[width=0.235\textwidth]{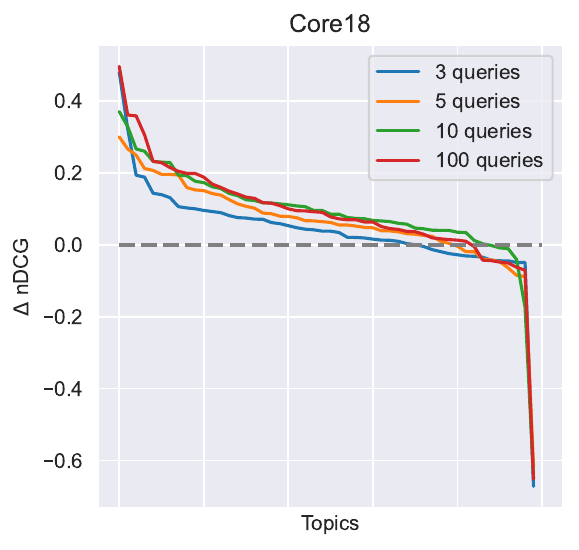}
  \includegraphics[width=0.235\textwidth]{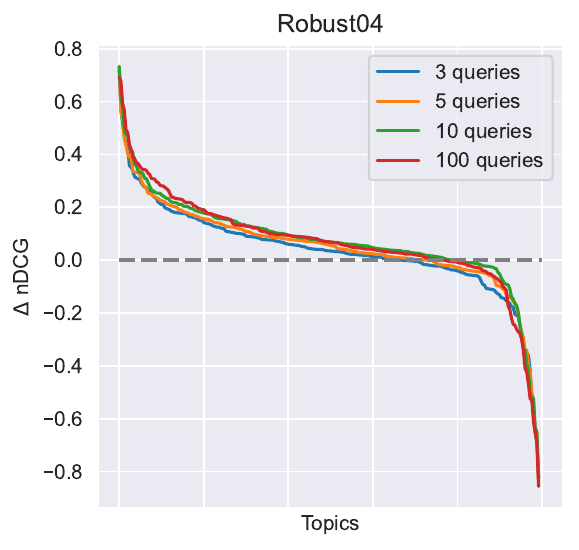}
  \includegraphics[width=0.235\textwidth]{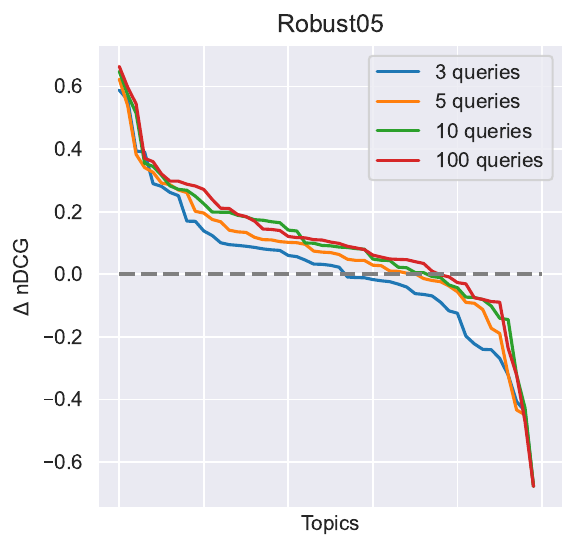}
  \caption{Retrieval effectiveness with different numbers of synthetic queries based on \textbf{P-2} \crule[black]{.3cm}{.3cm}\crule[cyan]{.3cm}{.3cm}\crule[purple]{.3cm}{.3cm}\crule[black]{.3cm}{.3cm}. The plots show the relative improvements in terms of $\Delta$ nDCG. For each topic of the fused rankings, the difference to the baseline (BM25 with the topic's title as the query) is determined with the four newswire benchmarks Core17/18 and Robust04/05.}
  \label{fig:delta_ndcg_num_queries}
\end{figure}

As generating synthetic query variants with \acp{LLM} is fairly cheap and unlimited in principle, an arbitrary number of queries for data fusion can be considered. To this end, Figure~\ref{fig:delta_ndcg_num_queries} compares the $\Delta \ \mathrm{nDCG}$ of the topic score distribution for fused rankings made with either 3, 5, 10, or 100 synthetic queries based on P-2. More specifically, we determine $\Delta \ \mathrm{nDCG}$ by the difference between the topic score of the fused ranking and that of the BM25 baseline ranking with a cutoff of 1,000 documents per ranking. Figure~\ref{fig:delta_ndcg_num_queries} shows these differences between the topic scores in decreasing order.

An increasing number of synthetic queries and fused rankings have a positive impact on the retrieval effectiveness. For instance, using more than three synthetic queries is always preferable for those topics that improve. However, the comparison of fused rankings based on ten and one hundred queries suggests that there are limits to this kind of improvement. Too many queries might cause topic drifts that eventually will not lead to any further improvements. In the future, a better understanding of how many synthetic queries are beneficial helps to leverage the full potential of the proposed method. As most of the $\Delta \ \mathrm{nDCG}$ scores are above the dashed line, fusing rankings generally positively impacts the effectiveness, as already concluded from the results in Table~\ref{tab:prompts.effectiveness}. However, there are some topics for which the retrieval effectiveness deteriorates, leaving room for further improvements.

\section{Discussion}

Overall, our experiments let us conclude that including the additional information, given by the topic description and narrative, in the prompt is beneficial for guiding the \ac{LLM} to produce topically related effective queries for data fusion, which provides several interesting directions for future work. 

\textbf{\textit{How can such a method be put into practice?}} In contrast to test collection-based retrieval experiments, real search sessions often do not have context-rich information as provided in the topic files. However, we envision a search application in which the users interact with a conversational search agent~\cite{DBLP:journals/ftir/ZamaniTDR23} to which the information need is expressed. Throughout the conversation, the search agent, i.e., the \ac{LLM}, would then generate query variants, relieving the users from the cognitive effort of formulating effective search queries. The fused rankings can then be shown directly to the users or serve as the retrieval component of a \ac{RAG} system~\cite{DBLP:conf/nips/LewisPPPKGKLYR020} to support the generated response with evidence. Furthermore, interactive conversations enable the integration of feedback loops with users and the agent taking turns. Users can then provide explicit feedback on the system outputs, helping the \ac{LLM} to adapt its query generation strategy throughout the conversation.

\textbf{\textit{Why not using the \ac{LLM} directly to rerank the documents?}} Earlier work suggests that \acp{LLM} can reliably identify relevant documents~\cite{DBLP:conf/sigir/0001SC024} and also rerank item lists for better retrieval effectiveness~\cite{DBLP:conf/sigir/MacAvaneyS23}. However, this is usually more costly, as the \ac{LLM} has to read more tokens than in our methodology. In addition to the prompt with the task and topic information, each document would need to be read in its entirety by the \ac{LLM}. Our approach uses the \ac{LLM} to generate short keyword queries, letting BM25 rank the documents with computationally lower costs. Depending on the \ac{LLM}, the length of input tokens is also a limitation, which can be critical for lengthy news articles like in our experimental setup. The document length does not matter for our proposed methodology. In this regard, future work should aim for a better understanding of the cost-effectiveness tradeoffs and also consider performance in general, as the traditional retrieval methods could provide efficiency benefits in interactive conversations with a user. For instance, comparing both computational and financial costs alongside retrieval performance (including effectiveness and efficiency) will give insights to leverage the proposed method optimally.

\textbf{\textit{Beyond data fusion}}, the provided query datasets could be exploited in user simulations. For example, the synthetic queries can serve as reformulations in a simulated search session. Future work should, therefore, aim to validate the fidelity wrt. real user query variants~\cite{DBLP:conf/ecir/BreuerFS22}. The dataset by Benham and Culpepper~\cite{DBLP:conf/adcs/BenhamC17} provides a perfect reference resource with real user queries for these endeavors. Similarly, these query datasets could also be used for analyzing alternative ways of pooling, where the focus shifts from system variability toward query variability~\cite{DBLP:conf/cikm/MoffatSTB15}, following the principle of query polyrepresentation~\cite{DBLP:journals/jasis/EfronW10}. From an ethical point of view, we note that any potential biases in the training data of the \ac{LLM} could also impact the generated queries. Future work should, therefore, not only compare the query variants to real queries but also evaluate if there are any biases in the query formulations, which would inevitably affect rankings and any other follow-up system outputs.

\section{Limitations}

Our experimental evaluations suggest that fusing rankings with generated query variants led to a decline in retrieval effectiveness for certain topics (cf. Figure~\ref{fig:delta_ndcg_num_queries}) if compared to a baseline with the topic's title as the query. A more in-depth understanding of the deteriorating topics could help to mitigate the loss of retrieval effectiveness and it is necessary to analyze for what kinds of topics and why the proposed methodology improves or deteriorates retrieval results. In this regard, potential topic drifts caused by the synthesized queries should also be considered. 

Even though the core methodology is straightforward from a high-level perspective, there is a huge amount of different methods, tools and resources available to implement it. For instance, other more open instruction-following \acp{LLM}~\cite{DBLP:conf/fat/LiesenfeldD24} could be leveraged for a more transparent and reproducible experimental setup. Similarly, other kinds of retrieval methods, like dense retrievers or different data fusion methods, should also be evaluated to gain better insights into the generalizability of the approach.
Smaller \acp{LLM} or models with lower quantization levels also offer promising ways to make the query generation possible in less hardware-demanding applications, and the effectiveness tradeoffs need to be evaluated. 

Likewise, there are also many opportunities for systematic analysis of the query generation that is based on prompting the \ac{LLM}. For instance, the temperature parameter, which was kept fixed in our experiments, could lead to different query outputs that very likely have an impact on the end-to-end retrieval performance. In our experiments, we guided the \ac{LLM} with real query variants from users, which did not yield better effectiveness than the prompts making use of the topic's title, description, and narrative. More in-depth analysis with more query logs is required to fully leverage the potential of keyword-based queries for \textit{chain-of-thought prompting}. Other (more detailed) prompting strategies could also be considered, and it should be evaluated how how well the prompting can instruct the model to produce the desired query outputs. 

\section{Conclusion}
\label{sec:discussion_conclusion}

Making use of query variants when searching for documents improves retrieval effectiveness, especially when it is operationalized with data fusion algorithms. This work analyzed if generative instruction-tuned \acp{LLM} can synthesize effective query variants for data fusion. In our experiments, instruction-tuned \acp{LLM} could indeed generate suitable query variants that significantly improve retrieval effectiveness when the \ac{LLM} is guided by the additional topic information obtained from the topic's description and narrative. We conclude that additional topical context information is key to fully leveraging the \ac{LLM}'s capability to generate effective queries for data fusion.

\section*{Acknowledgements}

We thank the anonymous reviewers for their valuable feedback on this work. This contribution has been developed in the project PLan\_CV. Within the funding programme FH-Personal, the project PLan\_CV (reference number 03FHP109) is funded by the German Federal Ministry of Education and Research (BMBF) and Joint Science Conference (GWK).

\balance
\bibliographystyle{ACM-Reference-Format}
\bibliography{references}

%%% -*-BibTeX-*-
%%% Do NOT edit. File created by BibTeX with style
%%% ACM-Reference-Format-Journals [18-Jan-2012].

\begin{thebibliography}{54}

%%% ====================================================================
%%% NOTE TO THE USER: you can override these defaults by providing
%%% customized versions of any of these macros before the \bibliography
%%% command.  Each of them MUST provide its own final punctuation,
%%% except for \shownote{}, \showDOI{}, and \showURL{}.  The latter two
%%% do not use final punctuation, in order to avoid confusing it with
%%% the Web address.
%%%
%%% To suppress output of a particular field, define its macro to expand
%%% to an empty string, or better, \unskip, like this:
%%%
%%% \newcommand{\showDOI}[1]{\unskip}   % LaTeX syntax
%%%
%%% \def \showDOI #1{\unskip}           % plain TeX syntax
%%%
%%% ====================================================================

\ifx \showCODEN    \undefined \def \showCODEN     #1{\unskip}     \fi
\ifx \showDOI      \undefined \def \showDOI       #1{#1}\fi
\ifx \showISBNx    \undefined \def \showISBNx     #1{\unskip}     \fi
\ifx \showISBNxiii \undefined \def \showISBNxiii  #1{\unskip}     \fi
\ifx \showISSN     \undefined \def \showISSN      #1{\unskip}     \fi
\ifx \showLCCN     \undefined \def \showLCCN      #1{\unskip}     \fi
\ifx \shownote     \undefined \def \shownote      #1{#1}          \fi
\ifx \showarticletitle \undefined \def \showarticletitle #1{#1}   \fi
\ifx \showURL      \undefined \def \showURL       {\relax}        \fi
% The following commands are used for tagged output and should be
% invisible to TeX
\providecommand\bibfield[2]{#2}
\providecommand\bibinfo[2]{#2}
\providecommand\natexlab[1]{#1}
\providecommand\showeprint[2][]{arXiv:#2}

\bibitem[Alaofi et~al\mbox{.}(2022)]%
        {DBLP:conf/sigir/AlaofiGMSSSSW22}
\bibfield{author}{\bibinfo{person}{Marwah Alaofi}, \bibinfo{person}{Luke Gallagher}, \bibinfo{person}{Dana McKay}, \bibinfo{person}{Lauren~L. Saling}, \bibinfo{person}{Mark Sanderson}, \bibinfo{person}{Falk Scholer}, \bibinfo{person}{Damiano Spina}, {and} \bibinfo{person}{Ryen~W. White}.} \bibinfo{year}{2022}\natexlab{}.
\newblock \showarticletitle{Where Do Queries Come From?}. In \bibinfo{booktitle}{\emph{{{SIGIR}}}}. \bibinfo{publisher}{{ACM}}, \bibinfo{pages}{2850--2862}.
\newblock


\bibitem[Alaofi et~al\mbox{.}(2023)]%
        {DBLP:conf/sigir/AlaofiGSS023}
\bibfield{author}{\bibinfo{person}{Marwah Alaofi}, \bibinfo{person}{Luke Gallagher}, \bibinfo{person}{Mark Sanderson}, \bibinfo{person}{Falk Scholer}, {and} \bibinfo{person}{Paul Thomas}.} \bibinfo{year}{2023}\natexlab{}.
\newblock \showarticletitle{Can Generative LLMs Create Query Variants for Test Collections? An Exploratory Study}. In \bibinfo{booktitle}{\emph{{SIGIR}}}. \bibinfo{publisher}{{ACM}}, \bibinfo{pages}{1869--1873}.
\newblock


\bibitem[Allan et~al\mbox{.}(2017)]%
        {DBLP:conf/trec/AllanHKLGV17}
\bibfield{author}{\bibinfo{person}{James Allan}, \bibinfo{person}{Donna Harman}, \bibinfo{person}{Evangelos Kanoulas}, \bibinfo{person}{Dan Li}, \bibinfo{person}{Christophe~Van Gysel}, {and} \bibinfo{person}{Ellen~M. Voorhees}.} \bibinfo{year}{2017}\natexlab{}.
\newblock \showarticletitle{{{TREC}} 2017 Common Core Track Overview}. In \bibinfo{booktitle}{\emph{{{TREC}}}} \emph{(\bibinfo{series}{{{NIST}} Special Publication}, Vol.~\bibinfo{volume}{500--324})}. \bibinfo{publisher}{{NIST}}.
\newblock


\bibitem[Aslam and Montague(2001)]%
        {DBLP:conf/sigir/AslamM01}
\bibfield{author}{\bibinfo{person}{Javed~A. Aslam} {and} \bibinfo{person}{Mark~H. Montague}.} \bibinfo{year}{2001}\natexlab{}.
\newblock \showarticletitle{Models for Metasearch}. In \bibinfo{booktitle}{\emph{{{SIGIR}}}}. \bibinfo{publisher}{{ACM}}, \bibinfo{pages}{275--284}.
\newblock


\bibitem[Azzopardi et~al\mbox{.}(2007)]%
        {DBLP:conf/sigir/AzzopardiRB07}
\bibfield{author}{\bibinfo{person}{Leif Azzopardi}, \bibinfo{person}{Maarten {de Rijke}}, {and} \bibinfo{person}{Krisztian Balog}.} \bibinfo{year}{2007}\natexlab{}.
\newblock \showarticletitle{Building Simulated Queries for Known-Item Topics: An Analysis Using Six European Languages}. In \bibinfo{booktitle}{\emph{{{SIGIR}}}}. \bibinfo{publisher}{{ACM}}, \bibinfo{pages}{455--462}.
\newblock


\bibitem[Bailey et~al\mbox{.}(2015)]%
        {DBLP:conf/sigir/BaileyMST15}
\bibfield{author}{\bibinfo{person}{Peter Bailey}, \bibinfo{person}{Alistair Moffat}, \bibinfo{person}{Falk Scholer}, {and} \bibinfo{person}{Paul Thomas}.} \bibinfo{year}{2015}\natexlab{}.
\newblock \showarticletitle{User Variability and {{IR}} System Evaluation}. In \bibinfo{booktitle}{\emph{{{SIGIR}}}}. \bibinfo{publisher}{{ACM}}, \bibinfo{pages}{625--634}.
\newblock


\bibitem[Bailey et~al\mbox{.}(2016)]%
        {DBLP:conf/sigir/BaileyMST16}
\bibfield{author}{\bibinfo{person}{Peter Bailey}, \bibinfo{person}{Alistair Moffat}, \bibinfo{person}{Falk Scholer}, {and} \bibinfo{person}{Paul Thomas}.} \bibinfo{year}{2016}\natexlab{}.
\newblock \showarticletitle{{{UQV100}}: {{A}} Test Collection with Query Variability}. In \bibinfo{booktitle}{\emph{{{SIGIR}}}}. \bibinfo{publisher}{{ACM}}, \bibinfo{pages}{725--728}.
\newblock


\bibitem[Baskaya et~al\mbox{.}(2013)]%
        {DBLP:conf/cikm/BaskayaKJ13}
\bibfield{author}{\bibinfo{person}{Feza Baskaya}, \bibinfo{person}{Heikki Keskustalo}, {and} \bibinfo{person}{Kalervo J{\"a}rvelin}.} \bibinfo{year}{2013}\natexlab{}.
\newblock \showarticletitle{Modeling Behavioral Factors in Interactive Information Retrieval}. In \bibinfo{booktitle}{\emph{{{CIKM}}}}. \bibinfo{publisher}{{ACM}}, \bibinfo{pages}{2297--2302}.
\newblock


\bibitem[Bassani(2022)]%
        {DBLP:conf/ecir/Bassani22}
\bibfield{author}{\bibinfo{person}{Elias Bassani}.} \bibinfo{year}{2022}\natexlab{}.
\newblock \showarticletitle{\texttt{ranx}: {{A}} Blazing-Fast Python Library for Ranking Evaluation and Comparison}. In \bibinfo{booktitle}{\emph{{{ECIR}} (2)}} \emph{(\bibinfo{series}{Lecture Notes in Computer Science}, Vol.~\bibinfo{volume}{13186})}. \bibinfo{publisher}{{Springer}}, \bibinfo{pages}{259--264}.
\newblock


\bibitem[Bassani and Romelli(2022)]%
        {DBLP:conf/cikm/BassaniR22}
\bibfield{author}{\bibinfo{person}{Elias Bassani} {and} \bibinfo{person}{Luca Romelli}.} \bibinfo{year}{2022}\natexlab{}.
\newblock \showarticletitle{\texttt{ranx.fuse}: {{A}} Python Library for Metasearch}. In \bibinfo{booktitle}{\emph{{{CIKM}}}}. \bibinfo{publisher}{{ACM}}, \bibinfo{pages}{4808--4812}.
\newblock


\bibitem[Belkin et~al\mbox{.}(1993)]%
        {DBLP:conf/sigir/BelkinCCC93}
\bibfield{author}{\bibinfo{person}{Nicholas~J. Belkin}, \bibinfo{person}{Colleen Cool}, \bibinfo{person}{W.~Bruce Croft}, {and} \bibinfo{person}{James~P. Callan}.} \bibinfo{year}{1993}\natexlab{}.
\newblock \showarticletitle{Effect of Multiple Query Representations on Information Retrieval System Performance}. In \bibinfo{booktitle}{\emph{{{SIGIR}}}}. \bibinfo{publisher}{{ACM}}, \bibinfo{pages}{339--346}.
\newblock


\bibitem[Belkin et~al\mbox{.}(1995)]%
        {DBLP:journals/ipm/BelkinKFS95}
\bibfield{author}{\bibinfo{person}{Nicholas~J. Belkin}, \bibinfo{person}{Paul~B. Kantor}, \bibinfo{person}{Edward~A. Fox}, {and} \bibinfo{person}{Joseph~A. Shaw}.} \bibinfo{year}{1995}\natexlab{}.
\newblock \showarticletitle{Combining the Evidence of Multiple Query Representations for Information Retrieval}.
\newblock  \bibinfo{volume}{31}, \bibinfo{number}{3} (\bibinfo{year}{1995}), \bibinfo{pages}{431--448}.
\newblock


\bibitem[Benham and Culpepper(2017)]%
        {DBLP:conf/adcs/BenhamC17}
\bibfield{author}{\bibinfo{person}{Rodger Benham} {and} \bibinfo{person}{J.~Shane Culpepper}.} \bibinfo{year}{2017}\natexlab{}.
\newblock \showarticletitle{Risk-Reward Trade-Offs in Rank Fusion}. In \bibinfo{booktitle}{\emph{{{ADCS}}}}. \bibinfo{publisher}{{ACM}}, \bibinfo{pages}{1:1--1:8}.
\newblock


\bibitem[Benham et~al\mbox{.}(2018a)]%
        {DBLP:conf/desires/BenhamCGLM18}
\bibfield{author}{\bibinfo{person}{Rodger Benham}, \bibinfo{person}{J.~Shane Culpepper}, \bibinfo{person}{Luke Gallagher}, \bibinfo{person}{Xiaolu Lu}, {and} \bibinfo{person}{Joel~M. Mackenzie}.} \bibinfo{year}{2018}\natexlab{a}.
\newblock \showarticletitle{Towards Efficient and Effective Query Variant Generation}. In \bibinfo{booktitle}{\emph{{{DESIRES}}}} \emph{(\bibinfo{series}{{{CEUR}} Workshop Proceedings}, Vol.~\bibinfo{volume}{2167})}. \bibinfo{publisher}{{CEUR-WS.org}}, \bibinfo{pages}{62--67}.
\newblock


\bibitem[Benham et~al\mbox{.}(2017)]%
        {DBLP:conf/trec/BenhamGMDCSMC17}
\bibfield{author}{\bibinfo{person}{Rodger Benham}, \bibinfo{person}{Luke Gallagher}, \bibinfo{person}{Joel~M. Mackenzie}, \bibinfo{person}{Tadele~Tedla Damessie}, \bibinfo{person}{Ruey-Cheng Chen}, \bibinfo{person}{Falk Scholer}, \bibinfo{person}{Alistair Moffat}, {and} \bibinfo{person}{J.~Shane Culpepper}.} \bibinfo{year}{2017}\natexlab{}.
\newblock \showarticletitle{{{RMIT}} at the 2017 {{TREC CORE}} Track}. In \bibinfo{booktitle}{\emph{{{TREC}}}} \emph{(\bibinfo{series}{{{NIST}} Special Publication}, Vol.~\bibinfo{volume}{500--324})}. \bibinfo{publisher}{{NIST}}.
\newblock


\bibitem[Benham et~al\mbox{.}(2018b)]%
        {DBLP:conf/trec/BenhamGML0SCM18}
\bibfield{author}{\bibinfo{person}{Rodger Benham}, \bibinfo{person}{Luke Gallagher}, \bibinfo{person}{Joel~M. Mackenzie}, \bibinfo{person}{Binsheng Liu}, \bibinfo{person}{Xiaolu Lu}, \bibinfo{person}{Falk Scholer}, \bibinfo{person}{J.~Shane Culpepper}, {and} \bibinfo{person}{Alistair Moffat}.} \bibinfo{year}{2018}\natexlab{b}.
\newblock \showarticletitle{{{RMIT}} at the 2018 {{TREC CORE}} Track}. In \bibinfo{booktitle}{\emph{{{TREC}}}} \emph{(\bibinfo{series}{{{NIST}} Special Publication}, Vol.~\bibinfo{volume}{500--331})}. \bibinfo{publisher}{{NIST}}.
\newblock


\bibitem[Benham et~al\mbox{.}(2019)]%
        {DBLP:journals/tois/BenhamMMC19}
\bibfield{author}{\bibinfo{person}{Rodger Benham}, \bibinfo{person}{Joel~M. Mackenzie}, \bibinfo{person}{Alistair Moffat}, {and} \bibinfo{person}{J.~Shane Culpepper}.} \bibinfo{year}{2019}\natexlab{}.
\newblock \showarticletitle{Boosting Search Performance Using Query Variations}.
\newblock \bibinfo{journal}{\emph{ACM Trans. Inf. Syst.}} \bibinfo{volume}{37}, \bibinfo{number}{4} (\bibinfo{year}{2019}), \bibinfo{pages}{41:1--41:25}.
\newblock


\bibitem[Breuer et~al\mbox{.}(2022)]%
        {DBLP:conf/ecir/BreuerFS22}
\bibfield{author}{\bibinfo{person}{Timo Breuer}, \bibinfo{person}{Norbert Fuhr}, {and} \bibinfo{person}{Philipp Schaer}.} \bibinfo{year}{2022}\natexlab{}.
\newblock \showarticletitle{Validating Simulations of User Query Variants}. In \bibinfo{booktitle}{\emph{{{ECIR}} (1)}} \emph{(\bibinfo{series}{Lecture Notes in Computer Science}, Vol.~\bibinfo{volume}{13185})}. \bibinfo{publisher}{{Springer}}, \bibinfo{pages}{80--94}.
\newblock


\bibitem[Breuer et~al\mbox{.}(2023)]%
        {DBLP:conf/jcdl/BreuerKST23}
\bibfield{author}{\bibinfo{person}{Timo Breuer}, \bibinfo{person}{Christin~Katharina Kreutz}, \bibinfo{person}{Philipp Schaer}, {and} \bibinfo{person}{Dirk Tunger}.} \bibinfo{year}{2023}\natexlab{}.
\newblock \showarticletitle{Bibliometric Data Fusion for Biomedical Information Retrieval}. In \bibinfo{booktitle}{\emph{{JCDL}}}. \bibinfo{publisher}{{IEEE}}, \bibinfo{pages}{107--118}.
\newblock


\bibitem[Brown et~al\mbox{.}(2020)]%
        {DBLP:conf/nips/BrownMRSKDNSSAA20}
\bibfield{author}{\bibinfo{person}{Tom~B. Brown}, \bibinfo{person}{Benjamin Mann}, \bibinfo{person}{Nick Ryder}, \bibinfo{person}{Melanie Subbiah}, \bibinfo{person}{Jared Kaplan}, \bibinfo{person}{Prafulla Dhariwal}, \bibinfo{person}{Arvind Neelakantan}, \bibinfo{person}{Pranav Shyam}, \bibinfo{person}{Girish Sastry}, \bibinfo{person}{Amanda Askell}, \bibinfo{person}{Sandhini Agarwal}, \bibinfo{person}{Ariel {Herbert-Voss}}, \bibinfo{person}{Gretchen Krueger}, \bibinfo{person}{Tom Henighan}, \bibinfo{person}{Rewon Child}, \bibinfo{person}{Aditya Ramesh}, \bibinfo{person}{Daniel~M. Ziegler}, \bibinfo{person}{Jeffrey Wu}, \bibinfo{person}{Clemens Winter}, \bibinfo{person}{Christopher Hesse}, \bibinfo{person}{Mark Chen}, \bibinfo{person}{Eric Sigler}, \bibinfo{person}{Mateusz Litwin}, \bibinfo{person}{Scott Gray}, \bibinfo{person}{Benjamin Chess}, \bibinfo{person}{Jack Clark}, \bibinfo{person}{Christopher Berner}, \bibinfo{person}{Sam McCandlish}, \bibinfo{person}{Alec Radford}, \bibinfo{person}{Ilya Sutskever}, {and} \bibinfo{person}{Dario Amodei}.} \bibinfo{year}{2020}\natexlab{}.
\newblock \showarticletitle{Language Models Are Few-Shot Learners}. In \bibinfo{booktitle}{\emph{{{NeurIPS}}}}.
\newblock


\bibitem[Chakraborty et~al\mbox{.}(2020)]%
        {DBLP:conf/cikm/ChakrabortyGC20}
\bibfield{author}{\bibinfo{person}{Anirban Chakraborty}, \bibinfo{person}{Debasis Ganguly}, {and} \bibinfo{person}{Owen Conlan}.} \bibinfo{year}{2020}\natexlab{}.
\newblock \showarticletitle{Retrievability Based Document Selection for Relevance Feedback with Automatically Generated Query Variants}. In \bibinfo{booktitle}{\emph{{{CIKM}}}}. \bibinfo{publisher}{{ACM}}, \bibinfo{pages}{125--134}.
\newblock


\bibitem[{Collins-Thompson} et~al\mbox{.}(2013)]%
        {DBLP:conf/trec/Collins-Thompson13}
\bibfield{author}{\bibinfo{person}{Kevyn {Collins-Thompson}}, \bibinfo{person}{Paul~N. Bennett}, \bibinfo{person}{Fernando Diaz}, \bibinfo{person}{Charlie Clarke}, {and} \bibinfo{person}{Ellen~M. Voorhees}.} \bibinfo{year}{2013}\natexlab{}.
\newblock \showarticletitle{{{TREC}} 2013 Web Track Overview}. In \bibinfo{booktitle}{\emph{{{TREC}}}} \emph{(\bibinfo{series}{{{NIST}} Special Publication}, Vol.~\bibinfo{volume}{500--302})}. \bibinfo{publisher}{{NIST}}.
\newblock


\bibitem[{Collins-Thompson} et~al\mbox{.}(2014)]%
        {DBLP:conf/trec/Collins-Thompson14}
\bibfield{author}{\bibinfo{person}{Kevyn {Collins-Thompson}}, \bibinfo{person}{Craig Macdonald}, \bibinfo{person}{Paul~N. Bennett}, \bibinfo{person}{Fernando Diaz}, {and} \bibinfo{person}{Ellen~M. Voorhees}.} \bibinfo{year}{2014}\natexlab{}.
\newblock \showarticletitle{{{TREC}} 2014 Web Track Overview}. In \bibinfo{booktitle}{\emph{{{TREC}}}} \emph{(\bibinfo{series}{{{NIST}} Special Publication}, Vol.~\bibinfo{volume}{500--308})}. \bibinfo{publisher}{{NIST}}.
\newblock


\bibitem[Cormack et~al\mbox{.}(2009)]%
        {DBLP:conf/sigir/CormackCB09}
\bibfield{author}{\bibinfo{person}{Gordon~V. Cormack}, \bibinfo{person}{Charles L.~A. Clarke}, {and} \bibinfo{person}{Stefan B{\"u}ttcher}.} \bibinfo{year}{2009}\natexlab{}.
\newblock \showarticletitle{Reciprocal Rank Fusion Outperforms Condorcet and Individual Rank Learning Methods}. In \bibinfo{booktitle}{\emph{{{SIGIR}}}}. \bibinfo{publisher}{{ACM}}, \bibinfo{pages}{758--759}.
\newblock


\bibitem[Dai et~al\mbox{.}(2023)]%
        {DBLP:conf/iclr/DaiZMLNLBGHC23}
\bibfield{author}{\bibinfo{person}{Zhuyun Dai}, \bibinfo{person}{Vincent~Y. Zhao}, \bibinfo{person}{Ji Ma}, \bibinfo{person}{Yi Luan}, \bibinfo{person}{Jianmo Ni}, \bibinfo{person}{Jing Lu}, \bibinfo{person}{Anton Bakalov}, \bibinfo{person}{Kelvin Guu}, \bibinfo{person}{Keith~B. Hall}, {and} \bibinfo{person}{Ming{-}Wei Chang}.} \bibinfo{year}{2023}\natexlab{}.
\newblock \showarticletitle{Promptagator: Few-shot Dense Retrieval From 8 Examples}. In \bibinfo{booktitle}{\emph{{ICLR}}}. \bibinfo{publisher}{OpenReview.net}.
\newblock


\bibitem[Efron and Winget(2010)]%
        {DBLP:journals/jasis/EfronW10}
\bibfield{author}{\bibinfo{person}{Miles Efron} {and} \bibinfo{person}{Megan~A. Winget}.} \bibinfo{year}{2010}\natexlab{}.
\newblock \showarticletitle{Query polyrepresentation for ranking retrieval systems without relevance judgments}.
\newblock \bibinfo{journal}{\emph{J. Assoc. Inf. Sci. Technol.}} \bibinfo{volume}{61}, \bibinfo{number}{6} (\bibinfo{year}{2010}), \bibinfo{pages}{1081--1091}.
\newblock


\bibitem[Gospodinov et~al\mbox{.}(2023)]%
        {DBLP:conf/ecir/GospodinovMM23}
\bibfield{author}{\bibinfo{person}{Mitko Gospodinov}, \bibinfo{person}{Sean MacAvaney}, {and} \bibinfo{person}{Craig Macdonald}.} \bibinfo{year}{2023}\natexlab{}.
\newblock \showarticletitle{{{Doc2Query{-}{-}}}: {{When}} Less Is More}. In \bibinfo{booktitle}{\emph{{{ECIR}} (2)}} \emph{(\bibinfo{series}{Lecture Notes in Computer Science}, Vol.~\bibinfo{volume}{13981})}. \bibinfo{publisher}{{Springer}}, \bibinfo{pages}{414--422}.
\newblock


\bibitem[Ingwersen(1996)]%
        {DBLP:journals/jd/Ingwersen96}
\bibfield{author}{\bibinfo{person}{Peter Ingwersen}.} \bibinfo{year}{1996}\natexlab{}.
\newblock \showarticletitle{Cognitive Perspectives of Information Retrieval Interaction: {{Elements}} of a Cognitive {{IR}} Theory}.
\newblock \bibinfo{journal}{\emph{J. Documentation}} \bibinfo{volume}{52}, \bibinfo{number}{1} (\bibinfo{year}{1996}), \bibinfo{pages}{3--50}.
\newblock


\bibitem[Jagerman et~al\mbox{.}(2023)]%
        {DBLP:journals/corr/abs-2305-03653}
\bibfield{author}{\bibinfo{person}{Rolf Jagerman}, \bibinfo{person}{Honglei Zhuang}, \bibinfo{person}{Zhen Qin}, \bibinfo{person}{Xuanhui Wang}, {and} \bibinfo{person}{Michael Bendersky}.} \bibinfo{year}{2023}\natexlab{}.
\newblock \showarticletitle{Query Expansion by Prompting Large Language Models}.
\newblock \bibinfo{journal}{\emph{CoRR}}  \bibinfo{volume}{abs/2305.03653} (\bibinfo{year}{2023}).
\newblock


\bibitem[Jordan et~al\mbox{.}(2006)]%
        {DBLP:conf/jcdl/JordanWG06}
\bibfield{author}{\bibinfo{person}{Chris Jordan}, \bibinfo{person}{Carolyn~R. Watters}, {and} \bibinfo{person}{Qigang Gao}.} \bibinfo{year}{2006}\natexlab{}.
\newblock \showarticletitle{Using Controlled Query Generation to Evaluate Blind Relevance Feedback Algorithms}. In \bibinfo{booktitle}{\emph{{{JCDL}}}}. \bibinfo{publisher}{{ACM}}, \bibinfo{pages}{286--295}.
\newblock


\bibitem[Khramtsova et~al\mbox{.}(2024)]%
        {DBLP:conf/sigir/KhramtsovaZBZ24}
\bibfield{author}{\bibinfo{person}{Ekaterina Khramtsova}, \bibinfo{person}{Shengyao Zhuang}, \bibinfo{person}{Mahsa Baktashmotlagh}, {and} \bibinfo{person}{Guido Zuccon}.} \bibinfo{year}{2024}\natexlab{}.
\newblock \showarticletitle{Leveraging LLMs for Unsupervised Dense Retriever Ranking}. In \bibinfo{booktitle}{\emph{{SIGIR}}}. \bibinfo{publisher}{{ACM}}, \bibinfo{pages}{1307--1317}.
\newblock


\bibitem[Larsen et~al\mbox{.}(2009)]%
        {DBLP:journals/jasis/LarsenIL09}
\bibfield{author}{\bibinfo{person}{Birger Larsen}, \bibinfo{person}{Peter Ingwersen}, {and} \bibinfo{person}{Berit Lund}.} \bibinfo{year}{2009}\natexlab{}.
\newblock \showarticletitle{Data fusion according to the principle of polyrepresentation}.
\newblock \bibinfo{journal}{\emph{J. Assoc. Inf. Sci. Technol.}} \bibinfo{volume}{60}, \bibinfo{number}{4} (\bibinfo{year}{2009}), \bibinfo{pages}{646--654}.
\newblock


\bibitem[Lewis et~al\mbox{.}(2020)]%
        {DBLP:conf/nips/LewisPPPKGKLYR020}
\bibfield{author}{\bibinfo{person}{Patrick S.~H. Lewis}, \bibinfo{person}{Ethan Perez}, \bibinfo{person}{Aleksandra Piktus}, \bibinfo{person}{Fabio Petroni}, \bibinfo{person}{Vladimir Karpukhin}, \bibinfo{person}{Naman Goyal}, \bibinfo{person}{Heinrich K{\"{u}}ttler}, \bibinfo{person}{Mike Lewis}, \bibinfo{person}{Wen{-}tau Yih}, \bibinfo{person}{Tim Rockt{\"{a}}schel}, \bibinfo{person}{Sebastian Riedel}, {and} \bibinfo{person}{Douwe Kiela}.} \bibinfo{year}{2020}\natexlab{}.
\newblock \showarticletitle{Retrieval-Augmented Generation for Knowledge-Intensive {NLP} Tasks}. In \bibinfo{booktitle}{\emph{NeurIPS}}.
\newblock


\bibitem[Liesenfeld and Dingemanse(2024)]%
        {DBLP:conf/fat/LiesenfeldD24}
\bibfield{author}{\bibinfo{person}{Andreas Liesenfeld} {and} \bibinfo{person}{Mark Dingemanse}.} \bibinfo{year}{2024}\natexlab{}.
\newblock \showarticletitle{Rethinking open source generative {AI:} open washing and the {EU} {AI} Act}. In \bibinfo{booktitle}{\emph{FAccT}}. \bibinfo{publisher}{{ACM}}, \bibinfo{pages}{1774--1787}.
\newblock


\bibitem[Lillis et~al\mbox{.}(2006)]%
        {DBLP:conf/sigir/LillisTCD06}
\bibfield{author}{\bibinfo{person}{David Lillis}, \bibinfo{person}{Fergus Toolan}, \bibinfo{person}{Rem~W. Collier}, {and} \bibinfo{person}{John Dunnion}.} \bibinfo{year}{2006}\natexlab{}.
\newblock \showarticletitle{{{ProbFuse}}: A Probabilistic Approach to Data Fusion}. In \bibinfo{booktitle}{\emph{{{SIGIR}}}}. \bibinfo{publisher}{{ACM}}, \bibinfo{pages}{139--146}.
\newblock


\bibitem[Lin et~al\mbox{.}(2021)]%
        {DBLP:journals/tois/LinYNTWL21}
\bibfield{author}{\bibinfo{person}{Sheng{-}Chieh Lin}, \bibinfo{person}{Jheng{-}Hong Yang}, \bibinfo{person}{Rodrigo~Frassetto Nogueira}, \bibinfo{person}{Ming{-}Feng Tsai}, \bibinfo{person}{Chuan{-}Ju Wang}, {and} \bibinfo{person}{Jimmy Lin}.} \bibinfo{year}{2021}\natexlab{}.
\newblock \showarticletitle{Multi-Stage Conversational Passage Retrieval: An Approach to Fusing Term Importance Estimation and Neural Query Rewriting}.
\newblock \bibinfo{journal}{\emph{{ACM} Trans. Inf. Syst.}} \bibinfo{volume}{39}, \bibinfo{number}{4} (\bibinfo{year}{2021}), \bibinfo{pages}{48:1--48:29}.
\newblock


\bibitem[MacAvaney and Soldaini(2023)]%
        {DBLP:conf/sigir/MacAvaneyS23}
\bibfield{author}{\bibinfo{person}{Sean MacAvaney} {and} \bibinfo{person}{Luca Soldaini}.} \bibinfo{year}{2023}\natexlab{}.
\newblock \showarticletitle{One-Shot Labeling for Automatic Relevance Estimation}. In \bibinfo{booktitle}{\emph{{SIGIR}}}. \bibinfo{publisher}{{ACM}}, \bibinfo{pages}{2230--2235}.
\newblock


\bibitem[MacAvaney et~al\mbox{.}(2021)]%
        {DBLP:conf/sigir/MacAvaneyYFDCG21}
\bibfield{author}{\bibinfo{person}{Sean MacAvaney}, \bibinfo{person}{Andrew Yates}, \bibinfo{person}{Sergey Feldman}, \bibinfo{person}{Doug Downey}, \bibinfo{person}{Arman Cohan}, {and} \bibinfo{person}{Nazli Goharian}.} \bibinfo{year}{2021}\natexlab{}.
\newblock \showarticletitle{Simplified Data Wrangling with \texttt{ir\_datasets}}. In \bibinfo{booktitle}{\emph{{{SIGIR}}}}. \bibinfo{publisher}{{ACM}}, \bibinfo{pages}{2429--2436}.
\newblock


\bibitem[Macdonald et~al\mbox{.}(2021)]%
        {DBLP:conf/cikm/MacdonaldTMO21}
\bibfield{author}{\bibinfo{person}{Craig Macdonald}, \bibinfo{person}{Nicola Tonellotto}, \bibinfo{person}{Sean MacAvaney}, {and} \bibinfo{person}{Iadh Ounis}.} \bibinfo{year}{2021}\natexlab{}.
\newblock \showarticletitle{{{PyTerrier}}: {{Declarative}} Experimentation in Python from {{BM25}} to Dense Retrieval}. In \bibinfo{booktitle}{\emph{{{CIKM}}}}. \bibinfo{publisher}{{ACM}}, \bibinfo{pages}{4526--4533}.
\newblock


\bibitem[Moffat et~al\mbox{.}(2015)]%
        {DBLP:conf/cikm/MoffatSTB15}
\bibfield{author}{\bibinfo{person}{Alistair Moffat}, \bibinfo{person}{Falk Scholer}, \bibinfo{person}{Paul Thomas}, {and} \bibinfo{person}{Peter Bailey}.} \bibinfo{year}{2015}\natexlab{}.
\newblock \showarticletitle{Pooled Evaluation over Query Variations: {{Users}} Are as Diverse as Systems}. In \bibinfo{booktitle}{\emph{{{CIKM}}}}. \bibinfo{publisher}{{ACM}}, \bibinfo{pages}{1759--1762}.
\newblock


\bibitem[Nogueira et~al\mbox{.}(2019)]%
        {DBLP:journals/corr/abs-1904-08375}
\bibfield{author}{\bibinfo{person}{Rodrigo~Frassetto Nogueira}, \bibinfo{person}{Wei Yang}, \bibinfo{person}{Jimmy Lin}, {and} \bibinfo{person}{Kyunghyun Cho}.} \bibinfo{year}{2019}\natexlab{}.
\newblock \showarticletitle{Document Expansion by Query Prediction}.
\newblock \bibinfo{journal}{\emph{CoRR}}  \bibinfo{volume}{abs/1904.08375} (\bibinfo{year}{2019}).
\newblock


\bibitem[Ounis et~al\mbox{.}(2005)]%
        {DBLP:conf/ecir/OunisAPHMJ05}
\bibfield{author}{\bibinfo{person}{Iadh Ounis}, \bibinfo{person}{Gianni Amati}, \bibinfo{person}{Vassilis Plachouras}, \bibinfo{person}{Ben He}, \bibinfo{person}{Craig Macdonald}, {and} \bibinfo{person}{Douglas Johnson}.} \bibinfo{year}{2005}\natexlab{}.
\newblock \showarticletitle{Terrier Information Retrieval Platform}. In \bibinfo{booktitle}{\emph{{{ECIR}}}} \emph{(\bibinfo{series}{Lecture Notes in Computer Science}, Vol.~\bibinfo{volume}{3408})}. \bibinfo{publisher}{{Springer}}, \bibinfo{pages}{517--519}.
\newblock


\bibitem[Penha et~al\mbox{.}(2022)]%
        {DBLP:conf/ecir/PenhaCH22}
\bibfield{author}{\bibinfo{person}{Gustavo Penha}, \bibinfo{person}{Arthur C{\^a}mara}, {and} \bibinfo{person}{Claudia Hauff}.} \bibinfo{year}{2022}\natexlab{}.
\newblock \showarticletitle{Evaluating the Robustness of Retrieval Pipelines with Query Variation Generators}. In \bibinfo{booktitle}{\emph{{{ECIR}} (1)}} \emph{(\bibinfo{series}{Lecture Notes in Computer Science}, Vol.~\bibinfo{volume}{13185})}. \bibinfo{publisher}{{Springer}}, \bibinfo{pages}{397--412}.
\newblock


\bibitem[Salemi et~al\mbox{.}(2024)]%
        {DBLP:conf/sigir/SalemiKZ24}
\bibfield{author}{\bibinfo{person}{Alireza Salemi}, \bibinfo{person}{Surya Kallumadi}, {and} \bibinfo{person}{Hamed Zamani}.} \bibinfo{year}{2024}\natexlab{}.
\newblock \showarticletitle{Optimization Methods for Personalizing Large Language Models through Retrieval Augmentation}. In \bibinfo{booktitle}{\emph{{SIGIR}}}. \bibinfo{publisher}{{ACM}}, \bibinfo{pages}{752--762}.
\newblock


\bibitem[Sanderson(2010)]%
        {DBLP:journals/ftir/Sanderson10}
\bibfield{author}{\bibinfo{person}{Mark Sanderson}.} \bibinfo{year}{2010}\natexlab{}.
\newblock \showarticletitle{Test Collection Based Evaluation of Information Retrieval Systems}.
\newblock \bibinfo{journal}{\emph{Found. Trends Inf. Retr.}} \bibinfo{volume}{4}, \bibinfo{number}{4} (\bibinfo{year}{2010}), \bibinfo{pages}{247--375}.
\newblock


\bibitem[Shokouhi(2007)]%
        {DBLP:conf/ecir/Shokouhi07a}
\bibfield{author}{\bibinfo{person}{Milad Shokouhi}.} \bibinfo{year}{2007}\natexlab{}.
\newblock \showarticletitle{Segmentation of Search Engine Results for Effective Data-Fusion}. In \bibinfo{booktitle}{\emph{{{ECIR}}}} \emph{(\bibinfo{series}{Lecture Notes in Computer Science}, Vol.~\bibinfo{volume}{4425})}. \bibinfo{publisher}{{Springer}}, \bibinfo{pages}{185--197}.
\newblock


\bibitem[Thomas et~al\mbox{.}(2024)]%
        {DBLP:conf/sigir/0001SC024}
\bibfield{author}{\bibinfo{person}{Paul Thomas}, \bibinfo{person}{Seth Spielman}, \bibinfo{person}{Nick Craswell}, {and} \bibinfo{person}{Bhaskar Mitra}.} \bibinfo{year}{2024}\natexlab{}.
\newblock \showarticletitle{Large Language Models can Accurately Predict Searcher Preferences}. In \bibinfo{booktitle}{\emph{{SIGIR}}}. \bibinfo{publisher}{{ACM}}, \bibinfo{pages}{1930--1940}.
\newblock


\bibitem[Voorhees(2004)]%
        {DBLP:conf/trec/Voorhees04b}
\bibfield{author}{\bibinfo{person}{Ellen~M. Voorhees}.} \bibinfo{year}{2004}\natexlab{}.
\newblock \showarticletitle{Overview of the {{TREC}} 2004 Robust Track}. In \bibinfo{booktitle}{\emph{{{TREC}}}} \emph{(\bibinfo{series}{{{NIST}} Special Publication}, Vol.~\bibinfo{volume}{500--261})}. \bibinfo{publisher}{{NIST}}.
\newblock


\bibitem[Voorhees(2006)]%
        {DBLP:journals/sigir/Voorhees06}
\bibfield{author}{\bibinfo{person}{Ellen~M. Voorhees}.} \bibinfo{year}{2006}\natexlab{}.
\newblock \showarticletitle{The {{TREC}} 2005 Robust Track}.
\newblock \bibinfo{journal}{\emph{SIGIR Forum}} \bibinfo{volume}{40}, \bibinfo{number}{1} (\bibinfo{year}{2006}), \bibinfo{pages}{41--48}.
\newblock


\bibitem[Voorhees and Ellis({[n.\,d.]})]%
        {DBLP:conf/trec/2018}
\bibfield{editor}{\bibinfo{person}{Ellen~M. Voorhees} {and} \bibinfo{person}{Angela Ellis}} (Eds.). \bibinfo{year}{[n.\,d.]}\natexlab{}.
\newblock \bibinfo{booktitle}{\emph{Proceedings of the Twenty-Seventh Text REtrieval Conference, {TREC} 2018, Gaithersburg, Maryland, USA, November 14-16, 2018}}. \bibinfo{series}{{NIST} Special Publication}, Vol.~\bibinfo{volume}{500-331}. \bibinfo{publisher}{NIST}.
\newblock


\bibitem[Voskarides et~al\mbox{.}(2020)]%
        {DBLP:conf/sigir/VoskaridesLRKR20}
\bibfield{author}{\bibinfo{person}{Nikos Voskarides}, \bibinfo{person}{Dan Li}, \bibinfo{person}{Pengjie Ren}, \bibinfo{person}{Evangelos Kanoulas}, {and} \bibinfo{person}{Maarten de Rijke}.} \bibinfo{year}{2020}\natexlab{}.
\newblock \showarticletitle{Query Resolution for Conversational Search with Limited Supervision}. In \bibinfo{booktitle}{\emph{{SIGIR}}}. \bibinfo{publisher}{{ACM}}, \bibinfo{pages}{921--930}.
\newblock


\bibitem[Wei et~al\mbox{.}(2022)]%
        {DBLP:journals/corr/abs-2201-11903}
\bibfield{author}{\bibinfo{person}{Jason Wei}, \bibinfo{person}{Xuezhi Wang}, \bibinfo{person}{Dale Schuurmans}, \bibinfo{person}{Maarten Bosma}, \bibinfo{person}{Ed~H. Chi}, \bibinfo{person}{Quoc Le}, {and} \bibinfo{person}{Denny Zhou}.} \bibinfo{year}{2022}\natexlab{}.
\newblock \showarticletitle{Chain-of-Thought Prompting Elicits Reasoning in Large Language Models}.
\newblock \bibinfo{journal}{\emph{CoRR}}  \bibinfo{volume}{abs/2201.11903} (\bibinfo{year}{2022}).
\newblock


\bibitem[Wu and Crestani(2002)]%
        {DBLP:conf/cikm/WuC02}
\bibfield{author}{\bibinfo{person}{Shengli Wu} {and} \bibinfo{person}{Fabio Crestani}.} \bibinfo{year}{2002}\natexlab{}.
\newblock \showarticletitle{Data Fusion with Estimated Weights}. In \bibinfo{booktitle}{\emph{{{CIKM}}}}. \bibinfo{publisher}{{ACM}}, \bibinfo{pages}{648--651}.
\newblock


\bibitem[Zamani et~al\mbox{.}(2023)]%
        {DBLP:journals/ftir/ZamaniTDR23}
\bibfield{author}{\bibinfo{person}{Hamed Zamani}, \bibinfo{person}{Johanne~R. Trippas}, \bibinfo{person}{Jeff Dalton}, {and} \bibinfo{person}{Filip Radlinski}.} \bibinfo{year}{2023}\natexlab{}.
\newblock \showarticletitle{Conversational Information Seeking}.
\newblock \bibinfo{journal}{\emph{Found. Trends Inf. Retr.}} \bibinfo{volume}{17}, \bibinfo{number}{3-4} (\bibinfo{year}{2023}), \bibinfo{pages}{244--456}.
\newblock


\end{thebibliography}

\end{document}